\documentclass[aps,prb,amsmath,amssymb,twocolumn,superscriptaddress,showpacs]{revtex4-2}

\usepackage{amsmath,amssymb}
\usepackage{graphicx}
\usepackage{appendix}
\usepackage{subcaption}
\usepackage[usenames,dvipsnames]{color}
\usepackage[colorlinks=true, citecolor=blue, linkcolor=blue, urlcolor=blue]{hyperref}
\usepackage{hyphenat}
\usepackage{txfonts}
\usepackage{dsfont}
\usepackage{enumitem}
\usepackage[utf8]{inputenc}

\begin{document}

\title{Ultimate resolution limits in coherent anti-Stokes Raman scattering imaging}
\author{Giacomo Sorelli}
\affiliation{Fraunhofer IOSB, Ettlingen, Fraunhofer Institute of Optronics,System Technologies and Image Exploitation, Gutleuthausstr. 1, 76275 Ettlingen, Germany}
\email{giacomo.sorelli@iosb.fraunhofer.de}
\author{Manuel Gessner}
\affiliation{Instituto de Física Corpuscular (IFIC), CSIC‐Universitat de València and Departament de Física Teòrica, UV, C/Dr Moliner 50, E-46100 Burjassot (Valencia), Spain}
\email{manuel.gessner@uv.es}
\author{Frank Schlawin}
\affiliation{Max Planck Institute for the Structure and Dynamics of Matter, Luruper Chaussee 149, 22761 Hamburg, Germany}
\affiliation{University of Hamburg, Luruper Chaussee 149, Hamburg, Germany}
\affiliation{The Hamburg Centre for Ultrafast Imaging, Luruper Chaussee 149, Hamburg D-22761, Germany}
\email{frank.schlawin@mpsd.mpg.de}

\begin{abstract}
Coherent anti-Stokes Raman scattering is a widely used imaging technique that provides chemical contrast without the need for labels, making it an extremely valuable tool in physics, chemistry, and biology. In this work, we explore its fundamental precision limits by applying tools from quantum information theory. We identify optimal measurement strategies and show that spatial mode demultiplexing--a technique already accessible in current experimental setups--can achieve these quantum limits and in many situations improve the sensitivity of conventional intensity measurements. Building on this, we introduce an advanced imaging scheme based on vortex beams, which we predict to enhance the image information in the final quantum state of light and thereby lead to even higher resolution and sensitivity. These findings establish a clear path for enhancing nonlinear imaging techniques using concepts from quantum science, bridging the gap between established microscopy methods and the emerging capabilities of quantum technologies.
\end{abstract}

\maketitle

\begin{figure*}
    \centering
    \includegraphics[trim={0 70 80 0}, clip, width=.9\textwidth]{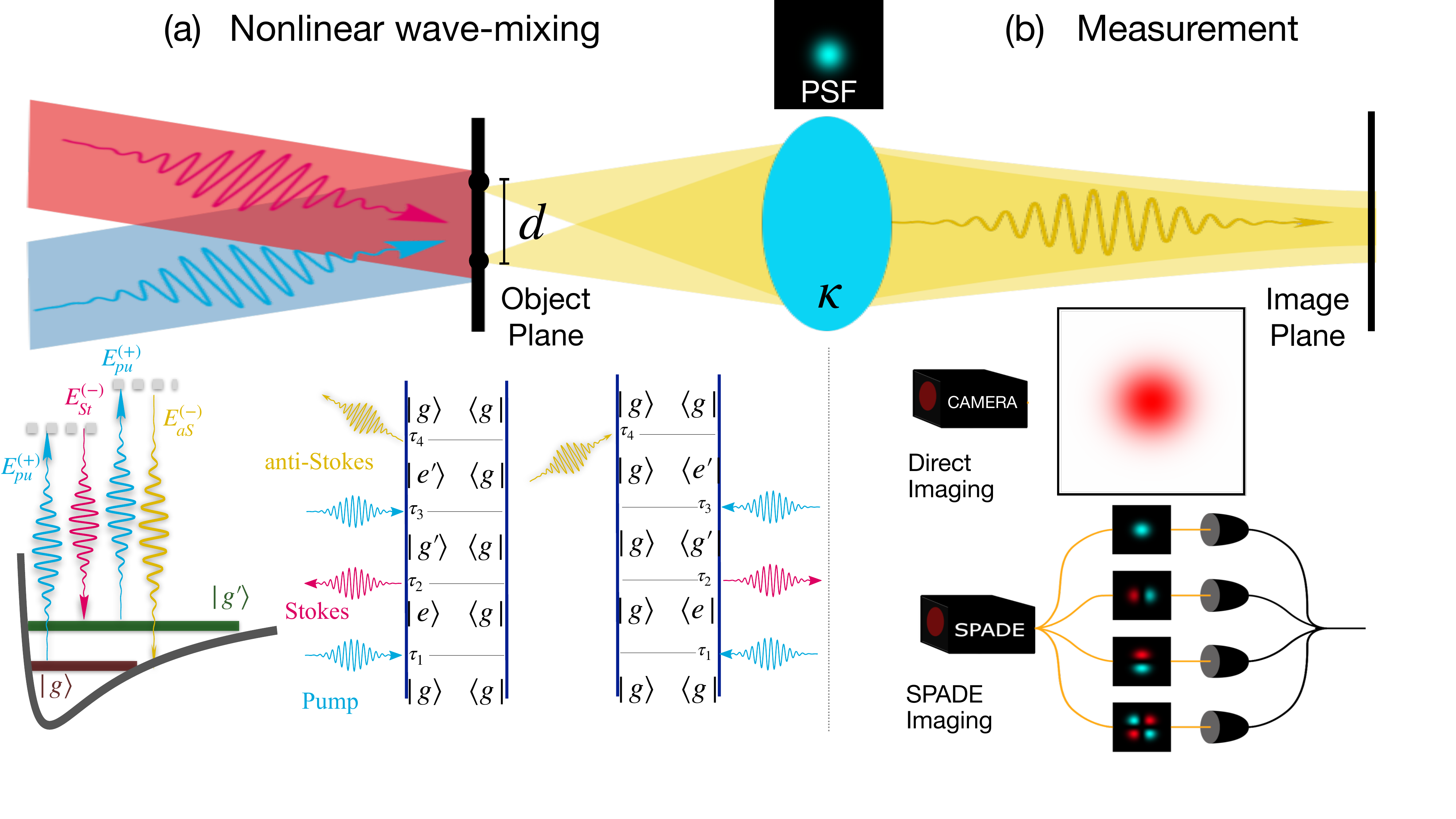}
    \caption{ 
    {\bf Sketch of the setup:} The two strong pump (blue) and Stokes (red) fields illuminate two point-like objects with distance $d$ in the object plane and stimulate emission of the corresponding anti-Stokes field (yellow). The emitted light is collected with a diffraction-limited imaging system with real symmetric point spread function (PSF) $u_0 (\vec{r})$ and transmission coefficient $\kappa$. 
    (a) CARS is a four-wave-mixing process, where excitation by pump and Stokes pulses, described by $\hat{E}_{\mathrm{pu}}$ and $\hat{E}_{\mathrm{St}}$, respectively, generates emission into the anti-Stokes sideband $\hat{E}_{\mathrm{aS}}$. The four-wave-mixing process is linked by double-sided Feynman diagrams to the microscopic dynamics in the molecule. The diagram rules for their evaluation are presented in the SM.
    (b) To estimate the source separation $d$, we consider different measurements in the image plane: (d) direct imaging, and (e) spatial mode demultiplexing (SPADE) in a PSF-adapted mode basis. 
    }
    \label{fig:CARS-diagrams}
\end{figure*}

\section{Introduction}
The resolution of optical microscopes fundamentally limits our ability to observe and characterize biological or chemical processes as well as 
our diagnostic capability in medical applications. 
It is therefore vital to derive fundamental resolution limits for imaging, in order to identify opportunities for technology advances and for the efficient implementation of quantum enhancements.
In linear imaging one performs a spatially resolved intensity measurement of the light collected by the microscope, e.g. with a camera. For this most ubiquitous technique, first steps towards were already taken. 
As already recognized by Abb\'e in the early days of microscopy, the resolution of this {\it direct imaging} approach is limited by diffraction \cite{Abbe1873}, or in other words it is impossible to resolve details with size smaller than the width of the point spread function (PSF) of the microscope, which is proportional to the ratio between the wavelength of the collected light and the numerical aperture of the microscope. 
Recently, Tsang {\it et al.} \cite{PhysRevX.6.031033} analysed this historical problem of estimating the separation between two incoherent point sources from the point of view of quantum metrology, and showed that spatial mode demultiplexing (SPADE) allows to resolve separations far below the diffraction limit. 
In fact, SPADE was proven to be the optimal imaging approach in the sense that it saturates the ultimate sensitivity limit imposed by quantum mechanics: the quantum Cram\'er-Rao bound \cite{helstrom1969,VittorioNP2011}.
The optimality of SPADE has been extended beyond incoherent sources \cite{Lupo2016,Tsang:19, Hradil:21, Kurdzialek2022, Sorelli_2022, Karuseichyk_2022,Karuseichyk_2024}, and proved to be robust against noise \cite{Len_2020,Lupo_2020,Gessner_2020,Almeida_2021,Sorelli_2021}.
Moreover, the superiority of SPADE over direct imaging for incoherent imaging has been verified in various proof of principle experiments \cite{Paur:16,Tham_2017,Boucher:20,Santamaria:23,Tan:23,Rouviere:24}.

These insights have opened up new pathways towards enhanced microscopy resolution inspired by quantum information techniques, but they are so far limited to linear techniques. Nonlinear imaging methods are highly relevant and widely used for studying biomedical systems~\cite{Parodi2020}, quantum materials~\cite{Pellatz2021, Chou2024} and molecular processes~\cite{Kukura2007, Batignani2024}. They encompass several incoherent and coherent techniques such as multi-photon excited fluorescence or higher harmonic generation. One of the most powerful nonlinear imaging techniques is coherent anti-Stokes Raman scattering (CARS) microscopy~\cite{Kukura2007, Zhang-review2018, Parodi2020}. It consists in a phase-sensitive four-wave mixing signal in which two strong pulses, pump and Stokes, stimulate emission of the anti-Stokes field~\cite{Rahav2010, Raman-book}, allowing for chemically sensitive, label-free microscopy~\cite{Kukura2007, Zhang-review2018, Parodi2020}.
In spite of the rapid progress in the improvement of this technique over recent years, important bottlenecks remain; in particular, the detection sensitivity and the resolution~\cite{Zhang-review2018}.
Current research focuses on the diffraction limit~\cite{Roadmap_superresolution, Tipping2024}, relying on the optimization of the excitation volume~\cite{Kim:12}, the possible use of vortex beams~\cite{Cho2024, Cho_vortex-beams}, strong illumination as in stimulated emission depletion~\cite{Silva2016, Gong2019} or phase-resolved signal extraction~\cite{DiazTormo:17, Zhitnitsky2024}. 
First groundbreaking experiments demonstrate the quantum-enabled enhancement of the signal-to-noise ratio of stimulated Raman imaging and spectroscopy, based on the photon counting of surface-enhanced signals~\cite{Meng2015, LiOpex2024, Li2025} or the use of nonclassical states of light~\cite{deAndrade:20, Casacio2021}. 
These advances demonstrate that quantum enhancements of these techniques are possible, but so far the fundamental resolution and sensitivity bounds for coherent Raman imaging remain unknown. Furthermore, a theoretical framework in which the potential enhancement by quantum technologies may be investigated systematically is still lacking.

Here, we provide such a formalism and use it to determine the quantum limits on resolution in CARS imaging. Our approach identifies feasible strategies to achieve quantum-enhanced imaging resolution and allows us to benchmark different measurement strategies systematically against the ultimate quantum limit. Specifically, we formulate CARS imaging of two molecular emitters as a parameter estimation problem and analyse it in the experimentally relevant regime of strong laser intensities. Our results show that subdiffraction imaging is feasible with the use of vortex light. Moreover, we propose spatial-mode demultiplexing as a means to enhanced signal strength and robustness.


\section{Results}

\textit{Fully quantum description of the CARS signal.--}Treatment of the CARS setup within the powerful framework of quantum metrology requires a fully quantum description of the process. Starting from the combined light-matter density matrix, we trace out the sample degrees of freedom to obtain a quantum map connecting the initial and final states of the light field, i.e.
\begin{align}
    \rho_{\mathrm{out}} &= \Phi [\rho_{\mathrm{in}}]. 
\end{align}
As shown in Ref.~\cite{arxiv2025}, $\Phi$ can be approximated in a cumulant expansion, $\Phi \simeq \exp [ \sum_n \mathcal{K}_n ]$, where the cumulants $\mathcal{K}_n$ can be identified with Feynman diagrams similar to those used in the semiclassical description of nonlinear spectroscopy. Here, they appear as superoperators acting on the photonic density matrix (see Methods). 

In the following, we analyse a conventional CARS signal generated by the excitation with strong, coherent pump and Stokes pulses. Mediated by a molecule at position $\vec{r}_1$, the wavemixing CARS four-wave-mixing process is given by the Feynman diagrams in Fig.~\ref{fig:CARS-diagrams}(a). We consider the initial state 
$\vert \psi_{\mathrm{in}} \rangle = \vert \alpha_{\mathrm{pu}} \rangle_{\mathrm{pu}} \otimes \vert \alpha_{\mathrm{St}}\rangle_{\mathrm{St}} \otimes \vert 0 \rangle_{\mathrm{aS}}$,
where $\alpha_{\mathrm{pu /pr}} \gg 1$. 
Neglecting the backaction of the CARS process on these strong fields, we replace the field operators of pump and Stokes fields by their amplitudes, i.e. $\hat{A}_{\mathrm{pu}} \rightarrow \alpha_{\mathrm{pu}}$ and $\hat{A}_{\mathrm{St}} \rightarrow \alpha_{\mathrm{St}}$, and restrict our explicit description to the anti-Stokes field. 
With the broadband field operator
$\hat{c}_{\mathrm{aS}} (\vec{r}) \equiv \int \frac{d\omega}{ 2\pi} \hat{a}_{\mathrm{aS}} (\vec{r}, \omega) \Phi (\omega)$  (see Methods), 
the full CARS signal of a molecule at position $\vec{r}_1$ evaluates to 
$\mathcal{K}_{\mathrm{CARS}} (\vec{r}_1) = \alpha (\vec{r}_1)\hat{c}_{aS, -} (\vec{r}_1)  - h.c.$, 
where the subscript "-" indicates the commutator superoperator, i.e. $c_{aS, -} \hat{X} = c_{\mathrm{aS}} \hat{X} - \hat{X} c_{\mathrm{aS}}$. 
Consequently, the molecules emits a coherent state, 
\begin{align} \label{eq.K_CARS}
    \rho (\vec{r}_1) &= \exp \left[ \mathcal{K}_{\mathrm{CARS}} (\vec{r}_1) \right] \vert 0 \rangle_{\mathrm{aS}} \langle 0 \vert_{\mathrm{aS}} = \vert \alpha (\vec{r}_1) \rangle \langle \alpha (\vec{r}_1) \vert,
\end{align}
with the coherent amplitude $\alpha (\vec{r}_1) = -i g u_{\mathrm{St}} (\vec{r}_1) \left( u^\ast_{\mathrm{pu}} (\vec{r}_1) \right)^2$.

\textit{CARS imaging.--}To analyse the resolution limit of CARS imaging, we consider the minimal problem already studied by Rayleigh \cite{Rayleigh}:
distinguishing two emitters in the object plane, defined by $z = 0$, at positions $\vec{r}_1$ and $\vec{r}_2$, respectively. 
According to Eq.~(\ref{eq.K_CARS}), the state emitted from the CARS process is a coherent state in two point-like modes at the position of the two molecules $\vert \alpha(\vec{r}_1), \alpha(\vec{r}_2) \rangle$. 

We observe the two emitters with a diffraction-limited imaging system with a (real-valued) PSF $u_0(\vec{r})$. Here we assume the quasi-monochromatic limit and paraxial propagation, such that the spectral mode $\Phi (\omega)$ is not distorted by the imaging system, though our theory could be amended when necessary.
The emitters' images 
are then broadened by the PSF, such that they
overlap by $\delta = \int d \vec{r}\; u_0(\vec{r} - \vec{r}_1) u_0(\vec{r} - \vec{r}_2)$. 
The final quantum state of the light field thus depends on two parameters: the coordinate of the centroid, i.e. $(\vec{r}_1 +\vec{r}_2)/2$, and the source separation $d = |\vec{r}_1 -\vec{r}_2|$. The imaging problem now consists in extracting maximal information about these parameters and the framework of quantum metrology allows us to identify the ultimate quantum limit as well as suitable observables and estimators for this task~\cite{Shapiro2009,Lupo2016,TsangReview,Sorelli_2021}. For simplicity, we will assume the centroid to be known, and treat CARS imaging as a separation estimation problem, leaving the more general case, where the centroid is 
estimated simultaneously with the separation to the Supplement (Appendix~\ref{appendix:QFIM-derivation}).



\textit{Quantum resolution limits.--}Within the parameter estimation framework, the resolution of the CARS imaging problem is given by the variance $(\Delta \tilde{d})^2$ of an unbiased estimator $\tilde{d}$ of the separation $d$.
According to the (quantum) Cramér-Rao lower bound, the variance $(\Delta \tilde{d})^2$ obeys the following chain of inequalities
\begin{equation}
    (\Delta \tilde{d})^2 \geq \frac{1}{\mu F_d} \geq \frac{1}{\mu Q_d},
\end{equation}
where $\mu$ is the total number of repeated measurements, $F_d$ is the Fisher information (FI) that bounds the estimation error when a specific measurement is performed, and
$Q_d$ is the quantum Fisher information (QFI) which is obtained from maximizing the FI over all possible observables allowed by quantum mechanics and thereby determines the ultimate sensitivity limit~\cite{BraunsteinPRL1994}. Accordingly, it only depends on the quantum state where the parameter is encoded \cite{helstrom1969}. 
A measurement that saturates the Cram\'er-Rao bound is guaranteed to be optimal and extract all the available information from the emitted light. Conversely, measurements that do not saturate the bound are suboptimal and miss out on potential sensitivity and/or resolution gains due to suboptimal data extraction.

For the problem at hand, the parameter of interest is not only encoded in the coherent states' amplitudes, but also in the shape of the modes, and both these dependencies need to be taken into account when computing the QFI. To account for this, we will make use of the framework of mode parameter estimation~\cite{Gessner2023,Sorelli2024} to determine the ultimate sensitivity for the estimation of the separation $d$ between two emitters from the CARS signal for arbitrary shapes of the excitation beams as well as the PSF of the microscope. Besides determining the ultimate resolution limit, we compare it with the achievable resolution of specific imaging modalities, focusing on direct imaging and spatial-mode demultiplexing (SPADE) followed by intensity measurements [see Fig.~\ref{fig:CARS-diagrams}(b)].
Direct imaging amounts to measuring the intensity distribution $I(\vec{r})$ in the image plane and represents the conventional approach. SPADE is an alternative technique that has proved to be particularly effective for separation estimation, both for coherent \cite{Tsang:19, Kurdzialek2022, Karuseichyk_2022} and incoherent sources \cite{PhysRevX.6.031033}, as well as in presence of separation-dependent coherence \cite{Karuseichyk_2024}.
\begin{figure}
\centering
\includegraphics[width =0.8\columnwidth]{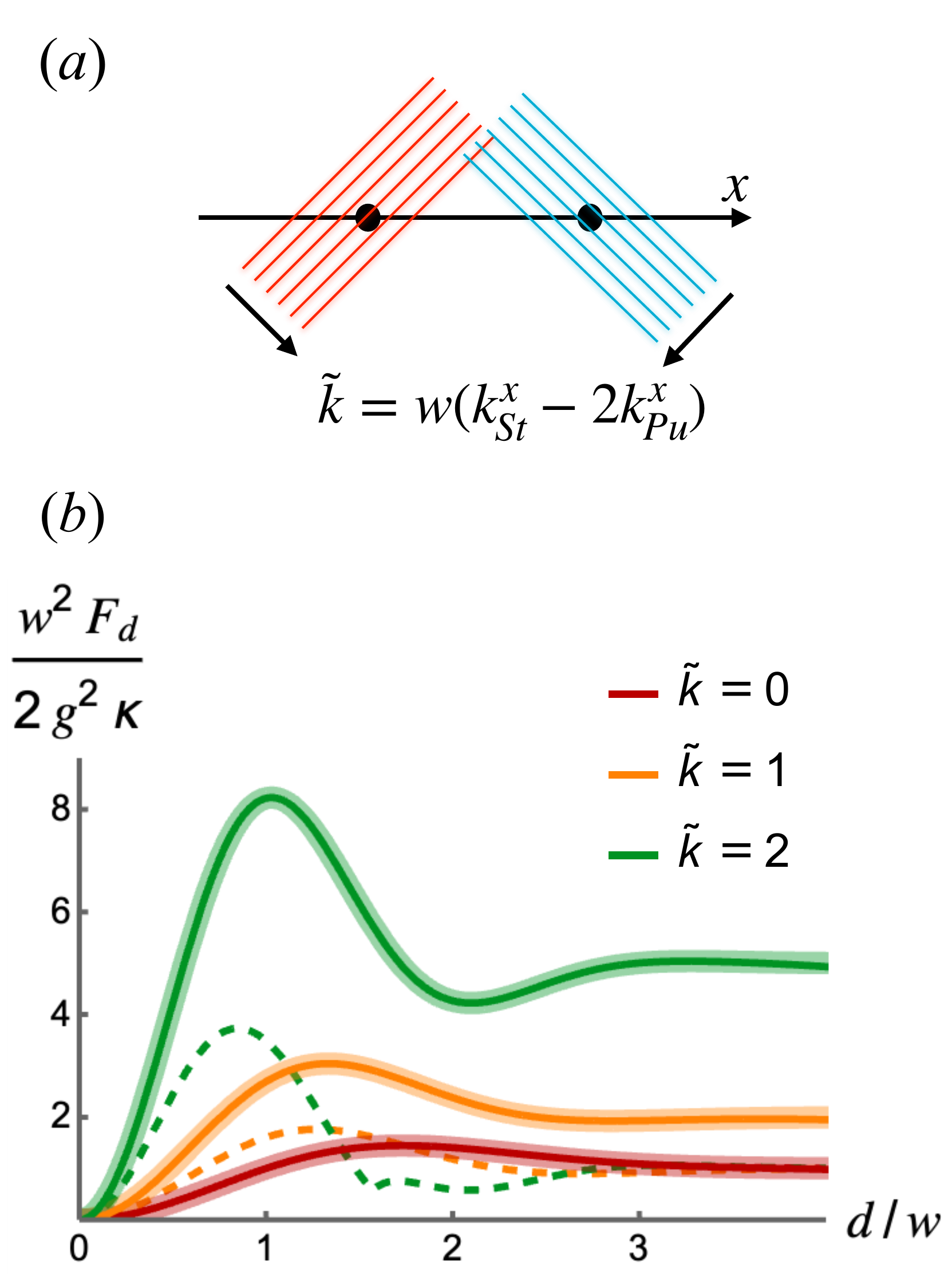}
\caption{{\bf CARS imaging with plane wave excitations} (a) Pictorial representation of two plane waves corresponding to the pump (red) and the Stokes (blue) pulses exciting two point emitters aligned along the $x$ axis (b) Fisher information $F_d$ as a function of the separation $d$ between the emitters (in units of the PSF width $w$), for different wave vectors of the excitation beams $\tilde{k} = w(k_{\mathrm{St}} - 2k_{\mathrm{pu}})$. 
Solid and dashed lines correspond to SPADE (with $M=10$ measured modes) and direct imaging, respectively. 
Broad bands corresponds to the quantum Fisher information $Q_d$.}
\label{Fig:FI-plane}
\end{figure}
Since SPADE depends on the shape of the PSF of the imaging system, we focus on a Gaussian PSF $u_0 (\vec{r})=(2/\pi w^2)^{1/4} \exp(- r^2/w^2)$, 
where the optimal SPADE basis is formed by Hermite-Gaussian (HG) modes $u_m(\vec{r})$ \cite{PhysRevX.6.031033, Rehacek:17}. Details on the derivation of the respective sensitivities are provided in the Methods section.

We gauge the resolution of the different strategies against the quantum limit using a plane wave approximation for the excitation pulses with spatial wave vectors $\vec{k}_{\mathrm{pu}}=(k^x_{\mathrm{pu}},k^y_{\mathrm{pu}} )$ and $\vec{k}_{\mathrm{St}} = (k^x_{\mathrm{St}},k^y_{\mathrm{St}} )$, for the pump and the Stokes beams, respectively. In this case, the dependence on the separation of the coherent state amplitude 
is modulated by the projection of the wave vectors of the excitation pulses along the axis of the sources $\tilde{k} = w(k^{x}_{\mathrm{St}} - 2k^{x}_{\mathrm{pu}})$. The corresponding behaviour of the QFI is compared to the FI for direct imaging and SPADE in Fig.~\ref{Fig:FI-plane}. 
This modulation can be used to significantly increase the estimation sensitivity, even for subdiffraction separations $d/w < 1$.
Direct imaging fails to fully exploit this increase: It only saturates the quantum Cram\'er-Rao bound for collinear excitation, $k^x_{\mathrm{St}} - 2k^x_{\mathrm{pu}} = 0$, i.e. when the coherent-state amplitudes $\alpha_\pm$ become independent on the source separation.
On the other hand, the FI for SPADE, represented as solid lines in Fig.~\ref{Fig:FI-plane} (b), always saturates the QFI.
Accordingly, SPADE allows to optimally extract information from the CARS signal and to reach the ultimate resolution limit.
Nevertheless, for $d\to 0$ the QFI vanishes, independently on the wave vector of the excitation beams. 
This is not surprising since two emitters with vanishingly small separation see the same excitation pulses, and therefore produce two in-phase coherent states that interfere constructively in the image plane. 
Such a constructive interference is known to produce a vanishing QFI \cite{Tsang:19, Hradil:21, Kurdzialek2022, Sorelli_2022, Karuseichyk_2022}. 
Note that here we have analysed a worst-case scenario of two identical emitters. 
In the presence of strong inhomogeneous broadening, this interference is disturbed, thus enabling a finite (albeit small) QFI even at vanishing distances. We analyse this situation in the Supplement. 

\begin{figure}
\centering
\includegraphics[width =\columnwidth]{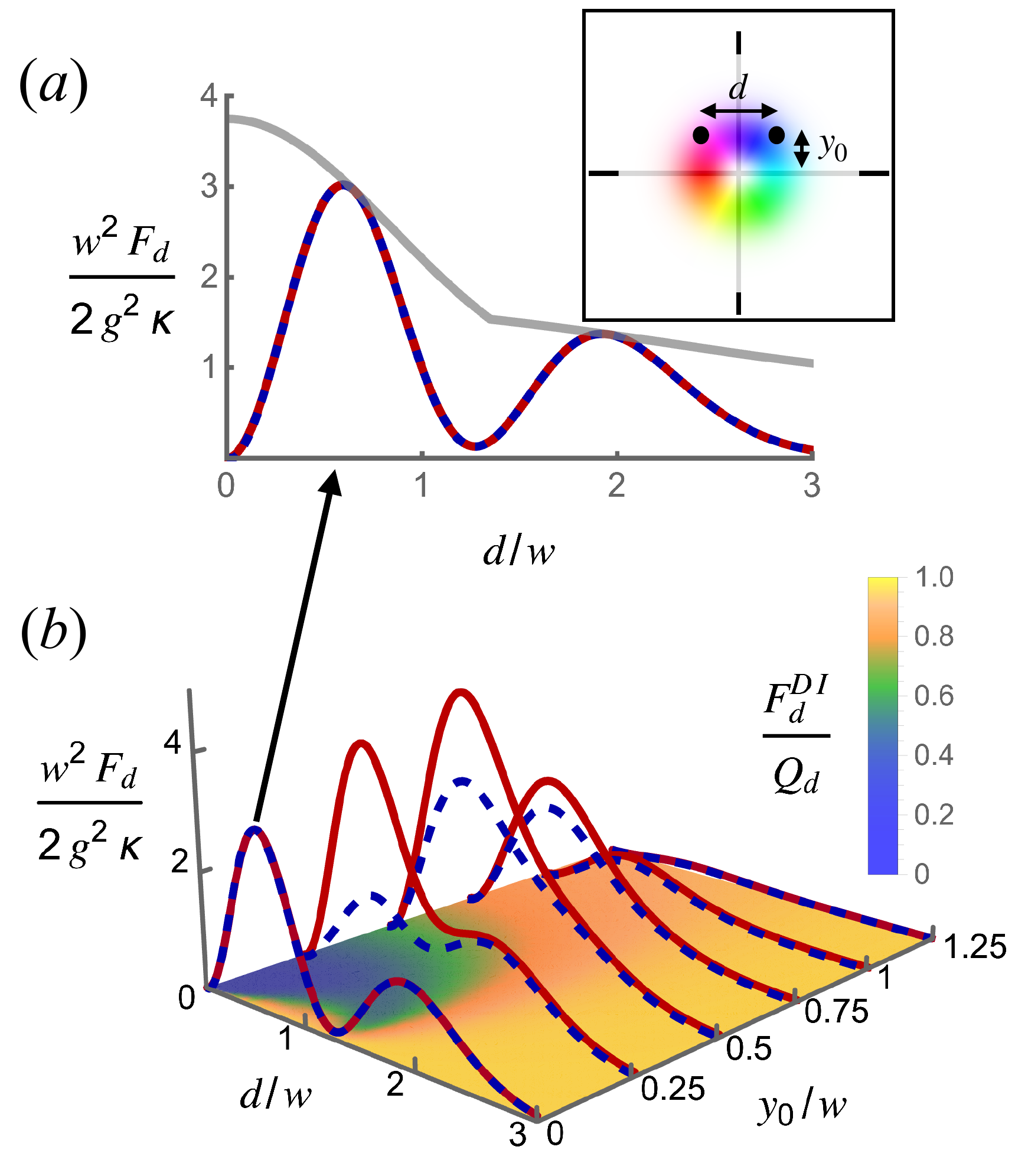}
\caption{{\bf CARS imaging with Laguerre-Gauss Stokes pulse} As illustrated in the black box in (a), two point emitters at distance $d$ are illuminated by opposite sides of the intensity ring of a Stokes pulse in a Laguerre Gauss mode (eventually shifted along the $y-$axis) superimposed with a spatially uniform pump pulse (not shown). (b) FI $F_d$ for SPADE (coinciding  with the QFI), $Q_d$ (solid red lines), and for direct imaging (dashed blue lines) as a function of the separation $d$ between the emitters (in units of the PSF width $w$), for different values of the vertical shift $y_0$ (in units of the PSF width $w$). 
The density plot at the bottom shows the ratio between the FI $F_d$ for direct imaging and the QFI, highlighting the region where SPADE provide an advantage.
All plots correspond to a ratio $a = w_{\mathrm{St}}/w = \sqrt{2}/2$ between the waist $w_{\mathrm{St}}$ of the Stokes pulse and the PSF width $w$. 
In (a) the curves for $y_0=0$ are repeated and compared with the QFI optimized over the excitation width $w_{\mathrm{St}}$ at every value of the separation $d$ (gray line).}
\label{Fig:FI-LG}
\end{figure}

The above discussion shows that  
it would be desirable to engineer the excitation such that the two sources emit out of phase. 
This can be achieved with a 
Stokes pulse in a Laguerre Gaussian (LG) mode $u_{\mathrm{St}}(\vec{r}) \propto r \exp(r^2/w_{\mathrm{St}}^2)\exp (i \phi)$, with $\phi$ the azimuthal angle in polar coordinates. 
The LG mode 
is characterized by a doughnut-shaped intensity distribution and a spiral phase, such that, according to Eq.~\eqref{eq.K_CARS}, if the two sources are excited by opposite sides of the doughnut they will emit out of phase, if  
the pump pulse features a flat phase. Here, we will consider a plane wave parallel to the source plane ($k^{Pu}_x = k^{Pu}_y = 0$). The technical details are presented in the Methods section.
If the excitation beam is slightly shifted by $y_0$ from the sources' axis, the two sources will emit with the same amplitude, but not exactly out of phase [see black box in Fig.~\ref{Fig:FI-LG} (a)]. 
The QFI achieved for this kind of excitation pulses is presented in red in Fig.~\ref{Fig:FI-LG} (b) for different values of the shift $y_0$, and for a fixed value of the ratio $a = w_{\mathrm{St}}/w$ between the waist $w_{\mathrm{St}}$ of the LG Stokes pulse and the width of the PSF $w$.
We observe a high sensitivity peak for subdiffraction separations, i.e. $d< w$, that can be further amplified by a small shift $y_0$.
Moreover, even though the QFI vanishes for $d\to 0$ for all values of $y_0$, by controlling the ratio $a$ one can push the high-sensitivity peak deeper into the subdiffraction region. 
This effect is illustrated by the gray line in Fig.~\ref{Fig:FI-LG} (a) which (for $y_0 = 0$) represents, for every value of the of the separation $d$, the QFI $Q_d$ maximized over $a$. 

At very small distances, such an effect requires subdiffraction phase gradients, that may be challenging to achieve with the LG beams that we chose here for analytical convenience. These are, however, not the only structured beams that could enhance spatial resolution in CARS imaging. For example, higher-order Bessel-Gaussian beams, that have a helical phase front and present multiple intensity rings with a radius smaller than their beam waist \cite{GORI_1987}, or super-oscillating light fields, that present features smaller than their wavelength \cite{Berry_2019}, could be used to induce the necessary phase gradient between emitters 
without the need for subdiffraction focusing.

Crucially, SPADE saturates the quantum Cram\'er-Rao bound for arbitrary values of the shift $y_0$. 
Direct imaging (dashed blue curves in Fig.~\ref{Fig:FI-LG}), in contrast, coincides with the QFI only for $y_0 = 0$, i.e. when the two sources emits perfectly out of phase. Even tiny displacements in $y$ substantially reduce the achievable sensitivity. 
Hence, this establishes SPADE as a much more robust way to extract the available image information and realize the full resolution limit of CARS imaging. 

\section{Discussion}
We have determined the ultimate quantum limits of resolution for CARS imaging by formulating the problem within the framework of quantum metrology. This perspective has identified potential improvements of existing setups by increasing the amount of relevant information that is encoded in the CARS signal and by extracting this information optimally.

Specifically, the quantum limit can be pushed towards higher resolution of two point emitters even at constant pulse intensity through the adequate design of the spatial beam profile of the excitation pulses. The more efficient information encoding is particularly important in situations where a small signal must be detected, as is the case, in particular, in the fingerprint region in complex biological environments. To get the most out of such low signals, it is not only important to encode information efficiently, but also to extract it optimally. Our results show that this is not always possible with standard direct imaging. Instead, SPADE detection can always saturate the ultimate sensitivity limit allowed by quantum mechanics.

Besides these results, 
our work opens new possibilities for nonlinear imaging in different contexts. 
Our technical derivation of the CARS signal based on Feynman diagrams and a cumulant expansion of the light-matter interaction Dyson series can be straightforwardly adapted to other imaging devices. 
Here, the CARS signal induced by strong laser pulses is then shown to be a coherent state, as one would intuitively expect from semiclassical theory. Based on the structure of the Feynman diagrams, this will also be true for any other nonlinear coherent imaging techniques, in the sense that the emitted light is coherent with distinct phase signature of the excitation pulses. Hence, we expect the conclusions reached in this study to apply more broadly. This concerns, for instance, the uncovered gains in resolution and signal strength that should also be interesting for second- and third-harmonic generation imaging. 

Beyond the coherent excitation scenario considered here, our approach further opens the possibility to systematically investigate the use of nonclassical states of light~\cite{Dorfman2016, Mukamel_2020} for nonlinear imaging. 
The use of these states as excitation pulses will break the symmetry of the Feynman diagrams and, in the case of CARS, give rise to incoherent and potentially nonclassical emission of the anti-Stokes fields. While this may seem undesirable at first glance, it might reveal new metrological opportunities to improve signal extraction~\cite{Panahiyan2022, Panahiyan2023} and to distinguish the sought-after signal from the incoherent background with suitable detection strategies: Since the incoherent background is produced by different wavemixing processes (see Supplement), the astute use of nonclassical states or spectral entanglement may enable the manipulation of the properties of the generated emission, and thereby render signal and background contributions distinguishable, or to even suppress the latter. 
In summary, our results lay the groundwork for the systematic application of optical quantum technologies in nonlinear imaging.

\section{Methods}

\textit{Cumulant expansion and CARS signal.--}Given the initial uncorrelated state $\varrho_0 = \rho_{\mathrm{in}} \otimes \rho_{\mathrm{sample}}$, where $\rho_{\mathrm{in}}$ is the initial state of the electromagnetic field and $\rho_{\mathrm{sample}}$ is the sample Hamiltonian, the combined light-matter density matrix after their interaction (i.e. $t \rightarrow \infty$) is given by
\begin{align}
    \varrho &= \mathcal{T} \exp \left[ - \frac{i}{\hbar} \int_{-\infty}^{\infty} \!\!\! d\tau\; \hat{H}_{I, -} (\tau)\right] \varrho_0,
\end{align}
where $\hat{H}_I (t)  = \hat{d}^{(-)} (\vec{r}, t) \hat{E}^{\dagger} (\vec{r}, t) + h.c.$ is the dipolar light-matter interaction Hamiltonian in the rotating wave approximation with the positive-frequency component of the dipole $\hat{d} (\vec{r},t)$ and positive-frequency component of the electric field operator $\hat{E} (\vec{r}, t)$ in the interaction picture. The subscript "-" denotes the commutator superoperator, i.e. $\hat{S}_- \hat{X} = \hat{S} \hat{X} - \hat{X} \hat{S}$. 

In the present case, we consider the interaction with three fields, i.e. $\hat{E} = \hat{E}_{\mathrm{pu}} + \hat{E}_{\mathrm{St}} + \hat{E}_{\mathrm{aS}}$. 
The pump and Stokes electric field operators are given by
\begin{align}
    \hat{E}_{\mathrm{pu}} (\vec{r}, t) &= E_0 u_{\mathrm{pu}} (\vec{r}) \psi_{\mathrm{pu}}(t) \hat{A}_{\mathrm{pu}}, \label{eq:E_pu} \\
    \hat{E}_{\mathrm{St}} (\vec{r}, t) &= E_0 u_{\mathrm{St}} (\vec{r}) \psi_{\mathrm{St}}(t) \hat{A}_{\mathrm{St}}, \label{eq:E_pr}
\end{align}
where $u_j (\vec{r})$ is the spatial and $\psi_j (t)$ is the temporal beam profile. The broadband operators $\hat{A}_j  = \int dt \; \psi^\ast_j (t) \hat{a} (t)$ inherit their bosonic nature from the temporal domain photon annihilation operators. In line with much of the theory on CARS spectroscopy, we here assume fixed polarization of the laser fields and do not treat them explicitly. 
The anti-Stokes field operator 
is written as
\begin{align}
    \hat{E}_{\mathrm{aS}} (\vec{r}, t) &= E_{\mathrm{aS}, 0} \int \frac{d\omega}{2\pi} \; \hat{a}_{\mathrm{aS}} (\vec{r}, \omega) e^{- i \omega t}, \label{eq:E_aS}
\end{align}
where $[\hat{a}_{\mathrm{aS}} (\vec{r}, \omega), \hat{a}_{\mathrm{aS}}^\dagger (\vec{r}', \omega')] = \delta^{(3)} (\vec{r}-\vec{r}') \delta (\omega - \omega')$ and $E_{aS, 0}$ is the vacuum field strength in the continuous frequency limit. 

Using the Feynman diagram rules~\cite{arxiv2025}, the diagrams in Fig.~\ref{fig:CARS-diagrams}(a) translate into the superoperator for a molecule located at position $\vec{r}_1$, we find
\begin{widetext}
\begin{align}
    \mathcal{K}_{\mathrm{CARS}} (\vec{r}_1) &= \left( - \frac{i}{\hbar} \right)^4 \int^\infty_{-\infty} \!\!\! d\tau_4 \int^{\tau_4} \!\!\! d\tau_3 \int^{\tau_3} \!\!\! d\tau_2 \int^{\tau_2} \!\!\! d\tau_1 \; \big\langle d^{(-)}_+ (\tau_4) d^{(+)}_L (\tau_3) d^{(-)}_L (\tau_2) d^{(+)}_L (\tau_1) \big\rangle \notag \\
    &\times \hat{E}_{\mathrm{aS}, -} (\vec{r}_1, \tau_4) \hat{E}_{\mathrm{pu}, L} (\vec{r}_1, \tau_3) \hat{E}^\dagger_{\mathrm{St}, L} (\vec{r}_1,\tau_2) \hat{E}_{\mathrm{pu}, L} (\vec{r}_1, \tau_1) + h.c., \label{eq:response-fct}
\end{align} 
where the subscripts "+", "L" and "R" indicate the anti-commutator, or action from left or right, respectively, $\hat{S}_+ \hat{X} = (\hat{S} \hat{X} + \hat{X} \hat{S}) / 2$, $\hat{S}_L \hat{X} = \hat{S} \hat{X}$, etc. 
Evaluating the response function in Eq.~(\ref{eq:response-fct}) [see Appendix~\ref{sec:CARS-derivation}] for a molecule initially in its ground state and the initial state $\vert \psi_{\mathrm{in}} \rangle $ of the main text, we obtain the superoperator for the anti-Stokes mode,
\begin{align} \label{eq.K_CARS1}
    \mathcal{K}_{\mathrm{CARS}} (\vec{r}_1) &= - i g u_{\mathrm{St}} (\vec{r}_1) \left( u^\ast_{\mathrm{pu}} (\vec{r})1) \right)^2\int \frac{d\omega}{2\pi} \; \hat{a}_{\mathrm{aS}, -}(\vec{r}_1,\omega) \Phi (\omega) + h.c.,
\end{align}
where we have split the CARS response into into a prefactor $g >0$ and a normalized spectral mode function $\Phi (\omega)$ with $\int d\omega / (2\pi) | \Phi (\omega) |^2 = 1$, i.e.
\begin{align}
    g \cdot \Phi (\omega) &\equiv \sum_{g'} \vert \alpha_{g', g}\vert^2 E_0^3 E_{\mathrm{aS}, 0} \int \frac{d\omega'}{2\pi} \int\frac{d\omega_-}{2\pi} \frac{\alpha_{\mathrm{pu}}^2 \alpha_{\mathrm{St}}}{\omega_- - \omega_{g'} + i \gamma_{g'}} \psi^\ast_{\mathrm{pu}} (\omega - \omega_-) \psi^\ast_{\mathrm{pu}} (\omega' + \omega_-) \psi_{\mathrm{St}} (\omega').
\end{align}
\end{widetext}
Here, $\alpha_{g'g}$ is the molecular polarizability, $\omega_{g'}$ the vibrational frequency, and $\gamma_{g'}$ its lifetime broadening, and $\alpha_{pu / St}$ the pump and Stokes amplitude.

In the imaging scenario discussed in the main text, the PSF $u_0 (\vec{r})$ maps the object plane state $\vert \alpha(\vec{r}_1), \alpha(\vec{r}_2) \rangle$ into non-orthogonal modes in the image plane. 
We can define two orthonormal modes of the two-emitter system based on the symmetric and antisymmetric superpositions of the two broadened images, $u_\pm (\vec{r}) = [u_0(\vec{r} - \vec{r_1}) \pm u_0(\vec{r} - \vec{r_2})] / (2(1\pm \delta))^{1/2}$. 
In this basis, the imaging system maps the object-plane coherent state $\vert \alpha(\vec{r}_1), \alpha(\vec{r}_2) \rangle$ into an image-plane coherent state $\vert\alpha_+, \alpha_-\rangle$ in the orthogonal modes with $\alpha_{\pm} = \sqrt{\kappa(1 \pm\delta)/2} [\alpha(\vec{r}_1)\pm \alpha(\vec{r}_2)]$,
where $\kappa$  is the transmission coefficient of the microscope.

\textit{(Quantum) Fisher for CARS imaging.--}Using the formalism for mode-encoded parameter estimation from Refs.~\cite{Gessner2023,Sorelli2024}, we obtain the following general expression for the QFI for the estimation of the separation $d$ between two emitters from the CARS signal (see App.~\ref{appendix:QFIM-derivation} for further details)
\begin{align}
    Q_d = 4\left[|\partial_d \alpha_+|^2 + |\partial_d \alpha_-|^2 + \eta_+^2 |\alpha_+|^2  + \eta_-^2 |\alpha_-|^2\right],
    \label{eq:QFI_d}
\end{align}
where $\partial_d$ denotes the partial derivative with respect to the separation 
$d$ and 
\begin{align}
    \eta_\pm^2 = \frac{(\Delta k)^2 \mp \beta}{4(1\pm \delta)} - \frac{(\delta^\prime)^2}{4(1\pm\delta^2)},
\end{align}
with $\delta^ \prime = \partial_d \delta$, $(\Delta k)^2 = \int   |\partial_x u_\pm (\vec{r})|^2 d \vec{r}$, and $\beta = \int \partial_x u_0{(\vec{r} - \vec{r_1})} \partial_x  u_0(\vec{r} - \vec{r_2}) d \vec{r}$.

We compare the QFI with the FI for specific measurements: direct imaging and SPADE.
The FI for the estimation of the separation $d$ from direct imaging can be written as \cite{PhysRevX.6.031033}
\begin{align}
   F_d^{\mathrm{DI}} = \int \!\! d\vec{r} \frac{1}{ I(\vec{r})}\left(\frac{\partial I(\vec{r})}{\partial d}\right)^ 2,
   \label{eq:FI_di}
\end{align}
where $I (\vec{r})$ is spatially resolved intensity distribution in the image plane.
In SPADE, the light in the image plane of the microscope is first decomposed into a basis of orthogonal spatial modes using a spatial light modulator. 
Then the number of photons are counted in each spatial component as illustrated in Fig.~\ref{fig:CARS-diagrams} (b).
The optimal mode basis for SPADE (in the sense of the quantum Cramér-Rao bound) can be constructed by orthogonalising successive derivatives of the PSF, e.g. for a Gaussian PSF this procedure leads to the HG modes \cite{Rehacek:17}. 
Experimental implementations of SPADE are typically based on multiplane light conversion \cite{Morizur:10}.
The FI for separation estimation from SPADE into $M$ modes can be written as \cite{PhysRevX.6.031033}
\begin{align}
   F_d^{\mathrm{SPADE}} = \sum_{m=0}^M  \frac{1}{N_m}\left(\frac{\partial N_m}{\partial d} \right)^2,
   \label{eq:FI-SPADE}
\end{align}
with $N_m$ the mean number of photons detected in mode $u_m(\vec{r})$.

\textit{CARS imaging with plane-wave excitation.--}When the two molecules are excited by plane waves $u_{\mathrm{pu}}(\vec{r}) = \exp (i \vec{k}_{\mathrm{pu}}\cdot \vec{r})$ and $u_{\mathrm{St}}(\vec{r}) = \exp (i \vec{k}_{\mathrm{St}}\cdot \vec{r})$, we obtain the following coherent-state amplitudes in the image plane
\begin{subequations}
\begin{align}
\alpha_+ &= -i g \sqrt{2\kappa(1+\delta)} \cos(\tilde{k} s/2) , \\
\alpha_- &= g \sqrt{2\kappa(1-\delta)} \sin (\tilde{k} s/2),
\end{align}
\label{eq:alpha_pm_plane_waves}
\end{subequations}
where we introduce $\tilde{k} = w (k^x_{\mathrm{St}} - 2k^x_{\mathrm{pu}})$, and $s = d/w$.
Substituting Eqs.~\eqref{eq:alpha_pm_plane_waves} into Eq.~\eqref{eq:QFI_d}, we obtain the QFI 
\begin{align}
    \frac{w^2 Q_d}{2\kappa g^2}  = 1+ \tilde{k}^2+ e^{-\frac{s^2}{2}} \left[\left(s^2-1-\tilde{k}^2\right) \cos (\tilde{k} s)+2 \tilde{k} s \sin (\tilde{k} s)\right].
    \label{eq:QFI_plane}
\end{align}
Equation~\eqref{eq:QFI_plane} is plotted as broad bands in Fig.~\ref{Fig:FI-plane} (b).

The spatially-resolved intensity distribution obtained from the CARS signal can be written as (see App. \ref{appendix:FI-derivation})
\begin{align}
    I(\vec{r}) = g^2\kappa &\left[u_0^2(\vec{r} -\vec{r}_1) + u_0^2(\vec{r} + \vec{r}_1) \right. \\
    &\left. + 2u_0(\vec{r} -\vec{r}_1)u_0(\vec{r} +\vec{r}_1)\cos (\tilde{k}s) \right], \nonumber
\end{align}
which can be used to compute the FI for direct imaging. 
The integral in Eq.~\eqref{eq:FI_di} cannot be taken analytically, but it can be easily evaluated numerically leading to the dashed curves in Fig.~\ref{Fig:FI-plane} (b).

When considering SPADE the mean number $N_m$ of photons detected in mode $u_m(\vec{r})$ can be written as (see App. \ref{appendix:FI-derivation})
\begin{align}
     N_m = 2\kappa g^2 \left[ 1 + (-1)^m \cos(\tilde{k}s)\right]\left(\frac{s}{2}\right)^{2m} \frac{e^{-s^2/2}}{m!}.
\end{align}
By substituting this expression into Eq.~\eqref{eq:FI-SPADE} for the FI and truncating the sum at $M = 10$ we obtain the solid lines in Fig.~\ref{Fig:FI-plane} (b), that for the range of considered separations, saturate the QFI. 

\textit{CARS imaging with a Laguerre-Gaussian Stokes pulse.--}We now consider the pump pulse to be in plane wave providing a uniform illumination of the source plane $u_{\mathrm{pu}}(\vec{r}) = \exp (i k_{\mathrm{pu}} z)$, and the Stokes pulse in the LG mode $u_{\mathrm{St}}(\vec{r}) = \mathcal{N}_{\mathrm{St}} r \exp (i\phi) \exp (-r^2/w_{\mathrm{St}}^2)$, where we choose the normalization constant $\mathcal{N}_{\mathrm{St}} = \sqrt{2 e}/w_{\mathrm{St}}$ such that the maximum of the intensity ring equals one.
We consider the possibility that the sources are shifted with respect to the optical axis of the Stokes pulse, i.e. they are located at $(\pm d/e, y_0$).
This results in the coherent-state amplitudes in the image plane
\begin{subequations}
\begin{align}
    \alpha_+ &= 2 g \sqrt{e} \sqrt{\kappa(1+\delta)}\frac{\psi}{a} \exp \left[-\left(\frac{s^2}{4} +\psi^2\right)/a^2\right],\\
    \alpha_- &= -i g \sqrt{e}\sqrt{\kappa(1-\delta)}\frac{s}{a}\exp \left[-\left(\frac{s^2}{4} +\psi^2\right)/a^2\right],
\end{align}
\label{eq:alpha_lg}
\end{subequations}
where we introduced $\psi = y_0/w$.
We can see that when the sources are aligned along the optical axis of the excitation beam (i.e. when $\psi = 0$), they emit completely out of phase and accordingly only the antisymmetric mode is populated, i.e $\alpha_+ = 0$. 
Substituting Eq.~\eqref{eq:alpha_lg} into Eq.~\eqref{eq:QFI_d}, we can obtain the QFI for arbitrary values of $\psi$. 
We present the full expression in App.~\ref{Appendix:particular_QFI}, and we report here only the particular case of $\psi=0$
\begin{align}
\label{eq:QFI-lg}
    \frac{w^2 Q_d}{2\kappa g^2}  &= \frac{e}{2a^6}e^{-s^2/2a^2}\left[s^4 +a^2s^2(a^2-4) +4a^2 \right. \\ 
    &\;+ \left. e^{-s^2/2} \left( (a^2-1)^2s^4+a^2(5a^2-4)s^2 + 4a^4\right)\right]. \nonumber
\end{align}
From Eq.~\eqref{eq:alpha_lg}, we can also evaluate the intensity distribution in the image plane for direct imaging, and the mean photon count per mode for SPADE. 
General results for arbitrary values of $\psi$ are reported in App.~\ref{appendix:FI-derivation}). 
The intensity distributions reads for $\psi = 0$
\begin{align}
    I(\vec{r}) = g^2\kappa e \left(\frac{s}{a}\right)^2 e^{-s^2/2a^2}\left[u_0(\vec{r} -\vec{r}_1) - u_0(\vec{r} + \vec{r}_1) \right]^2. 
\end{align}
By substituting this expression into Eq.~\eqref{eq:FI_di}, one can analytically verify that, in this case, the FI for direct imaging equals the QFI \eqref{eq:QFI-lg}.
This is however not true for $\psi \neq 0$ as illustrated in Fig.~\ref{Fig:FI-LG}.

In the case of SPADE measurements, we obtain the mean photon number in mode $m$
\begin{align}
    N_m = \frac{g^2 \kappa e}{a^2} \left(1 - (-1)^m\right)^2\frac{1}{m!}\left(\frac{s}{2}\right)^{2m+2}e^{-s^2/2}e^{-s^2/4a^2}.
\end{align}
If we substitute this expression into Eq.~\eqref{eq:FI-SPADE}, and consider SPADE in the full HG modes basis ($M\to \infty$), we can take the sum analytically, and show that FI for SPADE saturates the QFI \eqref{eq:QFI-lg}.
In contrast to direct imaging, the optimality of SPADE is preserved for all values of $\psi$ as illustrated in Fig.~\ref{Fig:FI-LG}.

\section{Acknowledgements}

We thank Dan Oron for helpful feedback on the manuscript. 
FS acknowledges support from the Cluster of Excellence 'Advanced Imaging of Matter' of the Deutsche Forschungsgemeinschaft (DFG) - EXC 2056 - project ID 390715994 and the research unit 'FOR5750: OPTIMAL' - project ID 531215165. This work is supported by the project PID2023-152724NA-I00, with funding from MCIU/AEI/10.13039/501100011033 and FSE+, by the project RYC2021-031094-I, with funding from MCIN/AEI/10.13039/501100011033 and the European Union ‘NextGenerationEU’ PRTR fund, by the project  CIPROM/2022/66 with funding by the Generalitat Valenciana, by the Ministry of Economic Affairs and Digital Transformation of the Spanish Government through the QUANTUM ENIA Project call—QUANTUM SPAIN Project, by the European Union through the Recovery, Transformation and Resilience Plan—NextGenerationEU within the framework of the Digital Spain 2026 Agenda, and by the CSIC Interdisciplinary Thematic Platform (PTI+) on Quantum Technologies (PTI-QTEP+). This work is supported through the project CEX2023-001292-S funded by MCIU/AEI and by the Fraunhofer Internal Programs under Grant No. Attract 40-09467.

\bibliography{bibliography_metrology}

\appendix
\renewcommand\thefigure{\thesection.\arabic{figure}}    
\setcounter{figure}{0}    

\begin{widetext}
\section{Derivation of Eq.~(\ref{eq.K_CARS1})}
\label{sec:CARS-derivation}

We consider a molecular sample model, as indicated in Fig.~\ref{fig:CARS-diagrams}(a), with a single ground state $\vert g\rangle$, electronically excited state $\vert e \rangle$ and vibrationally excited state $\vert g'\rangle$. They are coupled by the dipole operator
\begin{align}
    \hat{d} &= \mu_{ge} \vert e \rangle \langle g \vert + \mu_{g'e} \vert e \rangle \langle g' \vert + h.c.
\end{align}
Here, $\mu_{ij}$ denotes the matrix element of the dipole operator connecting levels $i$ and $j$. 
We further assume that the excited states are broadened by a loss mechanism described by decay rates $\gamma_{e}$ and $\gamma_{g'}$, respectively. 
The sample response function in the superoperator~(\ref{eq:response-fct}) can thus be evaluated straightforwardly,
\begin{align}
    \big\langle d^{(-)}_+ (\tau_4) d^{(+)}_L (\tau_3) d^{(-)}_L (\tau_2) d^{(+)}_L (\tau_1) \big\rangle &= \sum_{e, g'} \mu_{ge} \mu_{eg'} \mu_{g'e} \mu_{eg} e^{- i (\omega_{eg} - i \gamma_e) (\tau_2 - \tau_1)} e^{ - i (\omega_{e'g} - - \gamma_{g'}) (\tau_3 - \tau_2)} e^{- i (\omega_{e'g} - i\gamma_e) (\tau_4 - \tau_3)},
\end{align}
where $\hbar \omega_{ij}$ denotes the energy difference between states $i$ and $j$. 
In frequency domain, $\mathcal{K}_{\mathrm{CARS}}$ then reads, using Eqs.~(\ref{eq:E_pu})-(\ref{eq:E_aS})
\begin{align}
    \mathcal{K}_{\mathrm{CARS}} (\vec{r}_m) &= \left( - \frac{i}{\hbar}\right)^4 \sum_{e, g'} E_0^3 u_{\mathrm{St}} (\vec{r}) \left( u^\ast_{\mathrm{pu}} (\vec{r}) \right)^2 \int \frac{d\omega_1}{2\pi} \int \frac{d\omega_2}{2\pi} \int \frac{d\omega_3}{2\pi} \int \frac{d\omega_4}{2\pi} \; 2\pi \delta (\omega_1 - \omega_2 + \omega_3 - \omega_4) \notag \\
    &\times \frac{i \mu_{ge} }{\omega_1 - \omega_{eg} + i \gamma_e} \frac{i  \mu_{eg'} \mu_{g'e}  }{\omega_1 - \omega_2 - \omega_{g' g} + i \gamma_{g'}} \frac{i \mu_{eg} }{\omega_1 - \omega_2 + \omega_3 - \omega_{e' g} + i \gamma_e} \psi_{\mathrm{pu}}^\ast (\omega_1) \psi_{\mathrm{St}} (\omega_2) \psi_{\mathrm{pu}}^\ast (\omega_3) \notag \\
    &\times \hat{A}^2_{\mathrm{pu}, L} \hat{A}^\dagger_{\mathrm{St}, L} \hat{a}_{\mathrm{aS}, -} (\vec{r}_m, \omega_4) \label{eq:freq-domain-signal}
\end{align}
If the electronic excited states are far detuned from the frequencies of the involved light fields, we may replace the corresponding Green's functions by the detuning $\Delta$ between light and excitation, i.e.
\begin{align}
    \frac{i}{\omega_1 - \omega_{eg} + i \omega_e} \simeq \frac{i}{\omega_1 - \omega_2 + \omega_3 - \omega_{e' g} + i \gamma_e} \simeq \frac{i}{\Delta}. 
\end{align}
With the approximations described in the main text, and defining the molecular polarizability $\alpha_{g'g} = \mu_{ge} \mu_{g' e} / (\hbar^2 \Delta)$, 
we then arrive at Eq.~(\ref{eq.K_CARS1}).

We note that Eq.~(\ref{eq:freq-domain-signal}) on its own is insufficient to describe the evolution of the full density matrix comprising pump, Stokes, and anti-Stokes fields. Linear losses will affect the excitation pulses~\cite{Schlawin2025}, and other four-wave-mixing diagrams, including spontaneous and stimulated Raman gain and loss, but also distinct wavemixing processes such as two-photon absorption,
also need to take into account to properly describe changes to the excitation pulses. These may also become relevant when modelling the incoherent background in a particular experiment. Eq.~(\ref{eq:freq-domain-signal}) is, however, the only diagram leading to phase-dependent emission into the anti-Stokes mode, as described by Eq.~(\ref{eq.K_CARS1}).  

\section{QFI matrix for the joint estimation of separation and centroid} 
\label{appendix:QFIM-derivation}
In this appendix, we analyse CARS imaging as a two-parameter estimation problem. 
Without loss of generality, we assume that the two emitters lie on the $x-$axis of our coordinate system, so that the centroid reads $(\vec{r_1} +\vec{r_2})/2 = (x_0, 0)$. 
We consider the joint estimation of the parameters $\vec{\theta} = (d, x_0)$. 
The resolution of this two-parameter estimation problem is given by the covariance matrix $\Sigma_{kl} = Cov(\theta_k, \theta_l)$, with $\theta_1$ and $\theta_2$ locally unbiased estimators for the separation $d$ and the centroid $x_0$, respectively. 
According to the (quantum) Cram\'er Rao bound, the covariance matrix $\Sigma$ obeys the following chain of inequalities \cite{helstrom1969}
\begin{align}
    \Sigma \geq F^{-1}/\mu \geq Q^{-1}/\mu,
\end{align}
where $\mu$ is the total number of repeated measurements, $F$ is the Fisher information (FI) matrix, that bounds the estimation sensitivity when a specific measurement is performed, and $Q$ is the quantum Fisher information (QFI) matrix. The latter is obtained by maximizing the FI matrix over all possible observables allowed by quantum mechanics, accordingly it only depends on the the quantum state where the parameters are encoded \cite{helstrom1969}.

To compute the QFI matrix, we start from the CARS signal in the image plane $\vert \alpha_+, \alpha_-\rangle$, where $\alpha_\pm = \sqrt{\kappa(1 \pm\delta)/2} [\alpha(\vec{r}_1)\pm \alpha(\vec{r}_2)]$ are coherent-state amplitudes associated with the modes 
\begin{align}
    u_\pm (\vec{r}) = \frac{u_0(\vec{r} - \vec{r_1}) \pm u_0(\vec{r} - \vec{r_2})}{\sqrt{1\pm \delta}},
    \label{eq:upm_app}
\end{align}
and $\alpha(\vec{r})$ is determined by the pump and Stokes fields according to $\alpha(\vec{r}) = -i g u_{\mathrm{St}}(\vec{r}) \left(u_{\mathrm{pu}}^*(\vec{r}) \right)^2$. 
The coherent state $\vert \alpha_+, \alpha_-\rangle$ is a Gaussian state with unit covariance matrix, with the parameters $\vec{\theta}$ encoded both explicitly in the amplitudes $\alpha_\pm$ and implicitly in the shapes of the modes $u_\pm(\vec{r})$. 
Accordingly, we can compute the QFI matrix using a two-parameters generalization of the expression derived in \cite{Sorelli2024}, which, for coherent states, reads
\begin{align} 
Q_{k,l} &= 2 (\partial_{\theta_k} {\bf \bar{q}})^\top (\partial_{\theta_l} {\bf \bar{q}}) + 2 (\partial_{\theta_k} {\bf \bar{q}})^\top A_k {\bf \bar{q}} 
\\ &\; + 2 {\bf \bar{q}}^\top A_k^\top (\partial_{\theta_l} {\bf \bar{q}}) + 2{\bf \bar{q}}^\top(A_k^\top A_l + B_k^\top B_l ){\bf \bar{q}} \nonumber
\end{align}
where we introduced the mean field vector ${\bf \bar{q}} = 2(\Re[\alpha_+], \Im[\alpha_+],\Re[\alpha_-], \Im[\alpha_-])$. 
$A_k$ and $B_k$ are $4 \times 4$ matrices whose $nm =\pm$ blocks are given by 
\begin{subequations}
\begin{align}
    (A_k)_{nm} &= 
    \begin{pmatrix}
    \Re [a^k_{nm}] & - \Im [a^k_{nm}] \\ 
    \Im [a^k_{nm}] & \Re [a^k_{nm}] 
    \end{pmatrix},  \\
    (B_k)_{nm} &= 
    \begin{pmatrix}
    \Re [b^k_{nm}] & - \Im [b^k_{nm}] \\ 
    \Im [b^k_{nm}] & \Re [b^k_{nm}] 
    \end{pmatrix}, 
\end{align}
\label{eq:AB}
\end{subequations}
with 
\begin{subequations}
\begin{align} 
a^k_{nm} &= \int  u^*_m(\vec{r}) \partial_{\theta_k}u_n(\vec{r}) d \vec{r} \\ 
b^k_{nm} &= \int  \left(u^k_m(\vec{r})\right)^* u^k_n(\vec{r}) d \vec{r} ,
\end{align}
\label{eq:ab}
\end{subequations}
where $u^k_m(\vec{r})$ are obtained by orthogonalizing derivative modes $\partial_{\theta_k}u_n(\vec{r})$, which for the problem at hand read
\begin{subequations}
\begin{align}
u_\pm^{\theta_1=d} (\vec{r}) &= \partial_d u_\pm (\vec{r}),\\
u_\pm^{\theta_2=x_0} (\vec{r}) &=\partial_{x_0} u_\pm (\vec{r}) \pm \frac{\delta^\prime}{1-\delta^2}u_\mp (\vec{r}).
\end{align}
\label{eq:uk}
\end{subequations}

Substituting Eqs.~\eqref{eq:uk} and \eqref{eq:upm_app} into Eqs.~\eqref{eq:ab} and \eqref{eq:AB}, we obtain the following expressions for the matrices $A_k$ and $B_k$
\begin{subequations}
    \begin{align}
        A_1 &= 0_4, \\
        A_2 & = \begin{pmatrix}
            0_2 & \mathds{1}_2 \\
            -\mathds{1}_2 & 0_2
        \end{pmatrix},\\
        B_1 &= \eta_+ \mathds{1}_2\oplus \eta_-\mathds{1}_2, \\
        B_2 &= \xi_+ \mathds{1}_2\oplus \xi_-\mathds{1}_2,
    \end{align}
\end{subequations}
with 
\begin{subequations}
\begin{align}
    \eta_\pm^2 &= \int  |\partial_x u_\pm (\vec{r}) |^2 d \vec{r} = \frac{(\Delta k)^2 \mp \beta}{4(1\pm \delta)} - \frac{(\delta^\prime)^2}{4(1\pm\delta^2)}, \\
    \xi_\pm^2 &=  (\Delta k)^2 \mp \frac{\left(\delta^\prime\right)^2}{\delta(1-\delta^2)}\\
(\Delta k)^2 &= \int   |\partial_x u_\pm (\vec{r})|^2 d \vec{r},\\ 
\beta & = \int \partial_x u_0(\vec{r} - \vec{r_1}) \partial_x  u_0(\vec{r} - \vec{r_2}) d \vec{r}.
\end{align}
\label{eq:psf-related}
\end{subequations}
Putting all the pieces together, we  finally obtain the following expressions for the elements of the QFI matrix
\begin{subequations}
    \begin{align}
        Q_{11} = Q_d &=  4\left[|\partial_d \alpha_+|^2 + |\partial_d \alpha_-|^2 + \eta_+^2 |\alpha_+|^2  + \eta_-^2 |\alpha_-|^2\right] \\
        Q_{22} = Q_{x_0} &= 4 \left[|\partial_{x_0} \alpha_+|^2 + |\partial_{x_0} \alpha_-|^2 + \left(\xi_+^2 -\frac{\left(\delta^\prime\right)^2}{1-\delta^2}\right)|\alpha_+|^2  + \left(\xi_-^2 -\frac{\left(\delta^\prime\right)^2}{1-\delta^2}\right) |\alpha_-|^2\right] \\
        &\; + \frac{8\delta^\prime}{\sqrt{1-\delta^2}}\left( \Re\left[(\partial_{x_0} \alpha_+)^*\alpha_-\right] - \Re\left[(\partial_{x_0} \alpha_-)^*\alpha_+\right] \right) \nonumber\\
        Q_{12} = Q_{21} &= 4 \Re\left[(\partial_{d} \alpha_+)^*(\partial_{x_0}\alpha_+)\right] + 2 \Re\left[(\partial_{d} \alpha_-)^*(\partial_{x_0}\alpha_-)\right] + \frac{2\delta^\prime}{\sqrt{1-\delta^2}}\left(\Re\left[(\partial_{d} \alpha_+)^*\alpha_-\right] -  \Re\left[(\partial_{d} \alpha_-)^*\alpha_+\right]\right) \\
&\; + 2\eta_+\xi_+ |\alpha_+|^2  + 2 \eta_- \xi_- |\alpha_-|^2. \nonumber
    \end{align}
\end{subequations}
If we assume that the PSF of the microscope is given by a Gaussian $u_0 (\vec{r})=\sqrt{2/\pi w^2} \exp(- r^2/w^2)$, the quantities in Eqs.~\eqref{eq:psf-related} are given by
\begin{subequations}
\begin{align}
    \delta &= e^{-s^ 2/2}\\ 
    (\Delta k)^2 &= 1/w^2,\\ 
    \beta &=\frac{e^{-s^2/2}}{w^2}(s^ 2 -1)\\
    \eta_\pm^2 &= \pm\frac{1}{w^2}\frac{s^2 \pm \sinh\left(s^2/2\right)}{4(e^{s^ 2/4} \pm e^ {-s^2/4})^ 2}, \\
    \xi_\pm^2 &=  \frac{1}{w^2}\left(1 +\frac{s^2}{2\sinh(s^2/2)} \right), 
\end{align}
\label{eq_app:functions_gauss}
\end{subequations}
where we introduced the dimensionless variable $s = d/w$. 
The diagonal element $Q_d$ of the QFI matrix is reported as Eq~\eqref{eq:QFI_d} in the main text and is used to determine the ultimate sensitivity limit for separation estimation for different shapes of the excitation pulses.

\subsection{QFI for separation estimation with various excitation pulses}
\label{Appendix:particular_QFI}
\subsubsection{Plane-wave excitation pulses}
We now consider the Pump and Stokes beams to be plane waves, $u_{\mathrm{pu}}(\vec{r}) = \exp (i \vec{k}_{\mathrm{pu}}\cdot \vec{r})$ and $u_{\mathrm{St}}(\vec{r}) = \exp (i \vec{k}_{\mathrm{St}}\cdot \vec{r})$, respectively.
Since we assumed that the sources are aligned along the $x-$axis, we obtain the following expressions for the image plane coherent-state amplitudes
\begin{subequations}
\begin{align}
\alpha_+ &= -i g \sqrt{2\kappa(1+\delta)} \cos [(k^x_{\mathrm{St}} - 2k^x_{\mathrm{pu}}) d/2] , \\
\alpha_- &= g \sqrt{2\kappa(1-\delta)} \sin [(k^x_{\mathrm{St}} - 2k^x_{\mathrm{pu}}) d/2].
\end{align}
\label{eq_app:alpha_pm_plane_wave}
\end{subequations}

From Eqs.~\eqref{eq_app:alpha_pm_plane_wave} and \eqref{eq_app:functions_gauss}, we obtain the following expression for the QFI for separation estimation
\begin{align}
     \frac{w^2 Q_d}{2\kappa g^2}  = 1+ \tilde{k}^2+ e^{-\frac{s^2}{2}} \left[\left(s^2-1-\tilde{k}^2\right) \cos (\tilde{k} s)+2 \tilde{k} s \sin (\tilde{k} s)\right],
\end{align}
with $\tilde{k} = w (k^x_{\mathrm{St}} - 2k^x_{\mathrm{pu}})$.

\subsubsection{Laguerre-Gauss Stokes pulse}
\begin{figure}
    \centering
    \includegraphics[width=0.25\linewidth]{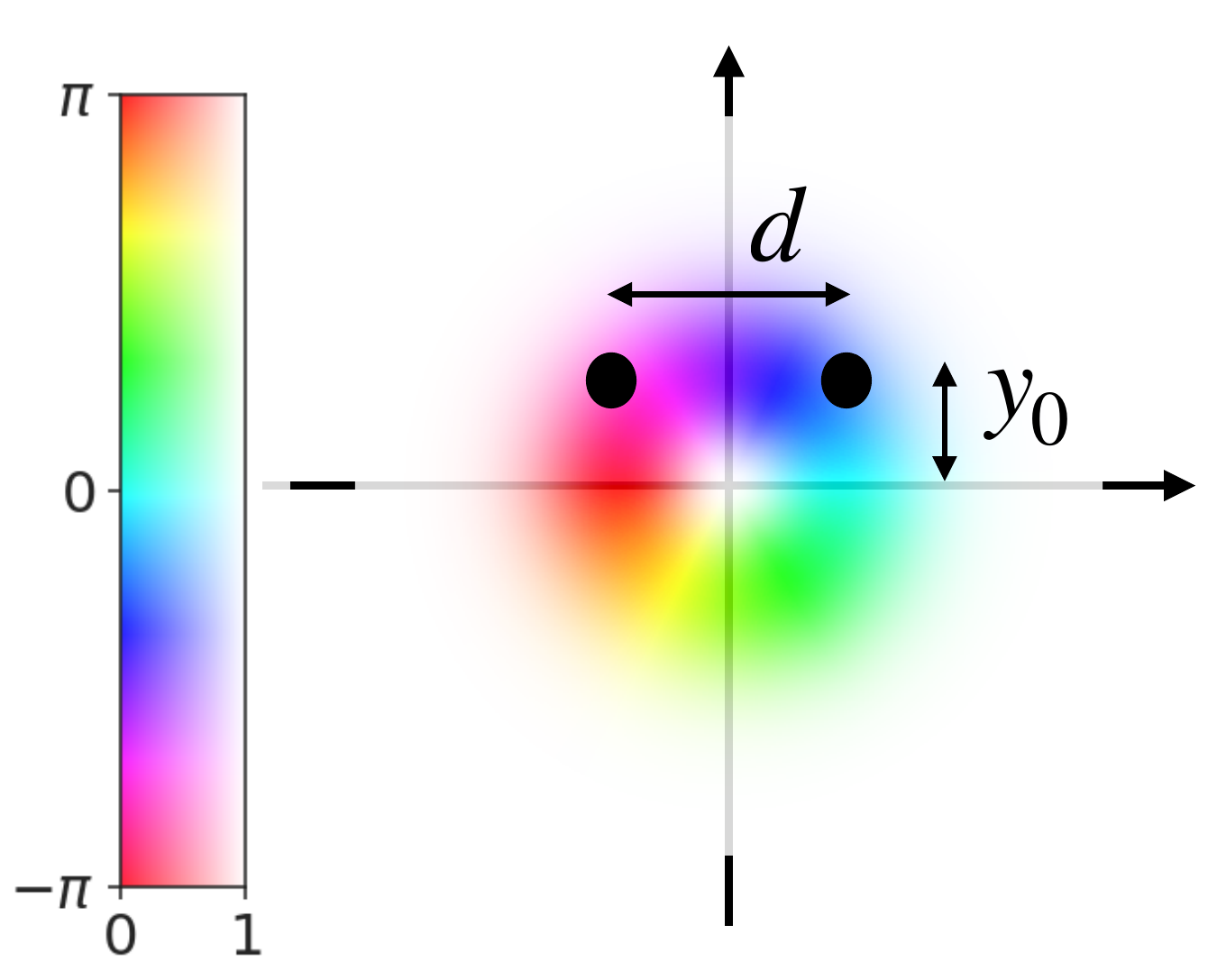}
    \caption{Laguerre Gaussian Stokes pulse exciting two sources not on axis.}
    \label{fig:shifted_excitation}
\end{figure}
Let us now consider the pump pulse in plane wave providing parallel to the source plane, $u_{\mathrm{pu}}(\vec{r}) = 1$, and the Stokes pulse in the LG mode $u_{\mathrm{St}}(\vec{r}) = \mathcal{N}_{\mathrm{St}} r \exp (i\phi) \exp (-r^2/w_{\mathrm{St}}^2)$, with the normalization constant $\mathcal{N}_{\mathrm{St}} = \sqrt{2 e}/w_{\mathrm{St}}$ such that the maximum of the intensity equals one.
We want to consider the possibility that the possibility of the optical axis of the Stokes pulse is not vertically aligned with the two emitters (see Fig.~\ref{fig:shifted_excitation}). 
To consider this case, it is more convenient to rewrite the LG mode $u_{\mathrm{St}}(\vec{r})$ in Cartesian coordinates
\begin{equation}
    u_{\mathrm{St}}(\vec{r}) =  \mathcal{N}_{\mathrm{St}} (x + i y) \exp \left[-(x^2 +y^2)/w_{\mathrm{St}}^2\right].
    \label{eq:LG_cart}
\end{equation}
If now assume the position of the sources to be $(\pm d/2, y_0)$, we obtain the coherent states amplitudes
\begin{subequations}
\begin{align}
    \alpha(\vec{r}_1) &= -i g \mathcal{N}_{\mathrm{St}} \left(-\frac{d}{2} + i y_0\right) \exp \left[-\left(\frac{d^2}{4} +y_0^2\right)/w_{\mathrm{St}}^2\right],\\
    \alpha(\vec{r}_2) &= -ig \mathcal{N}_{\mathrm{St}} \left(\frac{d}{2} + i y_0\right) \exp \left[-\left(\frac{d^2}{4} +y_0^2\right)/w_{\mathrm{St}}^2\right].
\end{align}  
\end{subequations}
Then using the mapping to the image plane, and introducing the dimensionless quantities $s = d/w$, $\psi = y_0/w$ and $a = w_{\mathrm{St}}/w$, we obtain the following expressions for the coherent-state amplitudes into the symmetric and antisymmetric modes 
\begin{subequations}
\begin{align}
    \alpha_+ &= 2 g \sqrt{e} \sqrt{\kappa(1+\delta)}\frac{\psi}{a} \exp \left[-\left(\frac{s^2}{4} +\psi^2\right)/a^2\right],\\
    \alpha_- &= -ig \sqrt{e} \sqrt{\kappa(1-\delta)}\frac{s}{a} \exp \left[-\left(\frac{s^2}{4} +\psi^2\right)/a^2\right].
\end{align}  
\label{eq:alphas_LG_y}
\end{subequations}
We can note that when the two sources are on opposite sides of the optical axis of the excitation beam ($\psi =0$), then they emit completely out of phase, and all the signal is the in the antisymmetric mode. 
On the other hand, if the the centroid of the two emitters doesn't lie on the optical axis of the Stokes pulse, then their phase difference will be smaller than $\pi$ and both the symmetric and the antisymmetric modes are populated. 
Substituting Eqs.~\eqref{eq:alphas_LG_y} into Eq.~\eqref{eq:QFI_d}, we obtain the QFI for separation estimation
\begin{align}
    Q_d = e \frac{e^{-\frac{s^2}{2a^2}}e^{-\frac{2\psi^2}{a^2}}}{2a^6}\left[s^4 +s^2 \left(4\psi^2 + a^2(a^2-4)\right) + 4a^4(1+\psi^2) - e^\frac{-s^2}{2} \left[s^4(a^2+1)^2 - s^2\left(a^2(5a^2+4) +4(a^2+1)^2\psi^2 \right) +4a^4(\psi^2+1)\right]\right],
\end{align}
which for sources with centroid on the optical axis of the Stokes pulse ($\psi=0$) reduces to
\begin{align}
    Q_d = e \frac{e^{-\frac{s^2}{2a^2}}}{2a^6}\left[s^4 +s^2 a^2(a^2-4) + 4a^4 - e^\frac{-s^2}{2} \left[s^4(a^2+1)^2 - s^2a^2(5a^2+4) +4a^4\right]\right].
\end{align}.

\section{FI for separation estimation and specific measurements}
\label{appendix:FI-derivation}
\subsection{Direct imaging}
Direct imaging is a spatially resolved intensity measurement in the image plane.
The electric field operator in the image plane can be written as $\hat{E}(\vec{r}) = \sum_{k=\pm} u_k(\vec{r})\hat{a}_k$, with $\hat{a}_\pm$ the annihilation operators in the symmetric and antisymmetric modes $u_\pm(\vec{r})$.
Accordingly, the mean intensity in the image plane can be written as 
\begin{align}
    I&(\vec{r}) = \langle\hat{E}^\dagger(\vec{r}) \hat{E}(\vec{r}) \rangle = \sum_{k,l=\pm} u_k(\vec{r}) u_l(\vec{r})\langle \hat{a}^\dagger_k\hat{a}_l \rangle = u^2_+(\vec{r}) |\alpha_+|^2 + u^2_-(\vec{r}) |\alpha_-|^2 + 2 \Re \left[\alpha_+^*\alpha_- \right] u_+(\vec{r})u_-(\vec{r})
\label{eq:intensity}
\end{align}
where we used that the CARS signal is a coherent state $|\alpha_+,\alpha_-\rangle$.
Since the coherent states $|\alpha_+, \alpha_-\rangle$ have Poissonian photon number statistics, we can write the FI for direct imaging as a simple function of mean intensity as \cite{kay1993}
\begin{align}
    F_d^{\mathrm{DI}} = \int d\vec{r} \frac{1}{ I(\vec{r})}\left(\frac{\partial I(\vec{r})}{\partial d}\right)^ 2.
    \label{eq:FI-DI-app}
\end{align}

 \subsubsection{Plane-wave excitation pulses}
 We now consider the Pump and Stokes beams to be the plane waves $u_{\mathrm{pu}}(\vec{r}) = \exp (i \vec{k}_{\mathrm{pu}}\cdot \vec{r})$ and $u_{\mathrm{St}}(\vec{r}) = \exp (i \vec{k}_{\mathrm{St}}\cdot \vec{r})$, respectively.
Substituting the expression for the coherent-state amplitudes $\alpha_\pm$ \eqref{eq_app:alpha_pm_plane_wave} into Eq.~\eqref{eq:intensity}, we obtain
\begin{align}
    I(\vec{r}) &= 2g^2\kappa\left[(1+\delta)u^2_+(\vec{r})\cos^2\left(\frac{\tilde{k} s}{2}\right) + (1-\delta)u^2_-(\vec{r})\sin^2\left(\frac{\tilde{k} s}{2}\right)\right] \\
    &=g^2\kappa \left[u_0^2(\vec{r} -\vec{r}_1) + u_0^2(\vec{r} + \vec{r}_1) + 2u_0(\vec{r} -\vec{r}_1)u_0(\vec{r} +\vec{r}_1)\cos (\tilde{k} s) \right], \nonumber
\end{align}
where we used the expression for the modes $u_\pm(\vec{r})$ \eqref{eq:upm_app}. For small separations, we can expand the intensity $I(\vec{r})$ to leading order in $s$, and compute the integral for the FI analytically. This yields
\begin{align}
     \frac{w^2 F_d^{\mathrm{DI}}}{2\kappa g^2} \sim \left(3+2\tilde{k}^2+\tilde{k}^4\right) s^2.
\end{align}
If we compare this with a leading order expansion of the QFI [Eq.~\eqref{eq:QFI_plane} in the main text]
\begin{align}
    \frac{w^2 Q_d}{2\kappa g^2} \sim \left(3+6\tilde{k}^2+\tilde{k}^4\right) s^2,
    \label{eq:QFI-small}
\end{align}
we see that while both expressions scale like $s^2$, the QFI coefficient is more favorable.
For arbitrary values of $s$ the integral in Eq.~\eqref{eq:FI-DI-app} cannot be taken analytically, but it can be easily evaluated numerically.

\subsubsection{Laguerre-Gauss Stokes pulse}
Let us consider the pump pulse in plane wave providing parallel to the source plane, $u_{\mathrm{pu}}(\vec{r}) = 1$, and the Stokes pulse in the LG mode $u_{\mathrm{St}}(\vec{r}) = \mathcal{N}_{\mathrm{St}} r \exp (i\phi) \exp (-r^2/w_{\mathrm{St}}^2)$.
Also in this case, we assume that the optical axis of the LG Stokes pulse might not be aligned with the centroid of the two emitters. 
Accordingly, combining Eq.~\eqref{eq:alphas_LG_y} with Eq.~\eqref{eq:intensity}, we obtain the intensity distribution 
\begin{align}
    I(\vec{r}) = g^2\kappa e \left(\frac{s}{a}\right)^2 e^{-(s^2/2 +\psi^2)/a^2}\left[u_0(\vec{r} -\vec{r}_1) - u_0(\vec{r} + \vec{r}_1) \right]^2 + g^2\kappa e \left(\frac{2\psi}{a}\right)^2 e^{-(s^2/2 +2\psi^2)/a^2}\left[u_0(\vec{r} -\vec{r}_1) +u_0(\vec{r} + \vec{r}_1) \right]^2.
\end{align}
For general values of $\psi$, the integral in Eq.~\eqref{eq:FI-DI-app} cannot be taken analytically, but it can be easily evaluated numerically. 
In general this yields a value of the FI which is lower than the QFI.
For the particular case of $\psi=0$, we can take the integral in Eq.~\eqref{eq:FI-DI-app} analytically, and we obtain the FI 
\begin{align}
\label{eq:FI-di-lg}
    \frac{w^2 F^{\mathrm{DI}}_d}{2\kappa g^2}  &= \frac{e}{2a^6}e^{-s^2/2a^2}\left[s^4 +a^2s^2(a^2-4) +4a^2 + e^{-s^2/2} \left( (a^2-1)^2s^4+a^2(5a^2-4)s^2 + 4a^4\right)\right],
\end{align}
which coincides with the QFI~\eqref{eq:QFI-lg} in the main text.

\subsection{Spatial-mode demultiplexing}
Spatial-mode demultiplexing (SPADE) consists of photon counting measurements in a set of PSF-adapted modes. 
In particular, for a Gaussian PSF $u_0 (\vec{r})=\sqrt{2/\pi w^2} \exp(- r^2/w^2)$, these are the Hermite-Gauss (HG) modes
\begin{align}
    u_{k} (\vec{r}) = \frac{1}{\sqrt{2^k k!}}H_n\left(\frac{\sqrt{2}x}{w} \right)u_0(r),
\end{align}
with $H_n(x)$ are Hermite polynomials. 
For the sake of simplicity, we are considering here SPADE aligned with the sources centroid, which we assume to be the origin of our coordinate system so that we have $x_0 =0$, $\vec{r}_1 = (-d/2,0)$ and $\vec{r}_2 = (d/2,0)$.

To describe SPADE it is convenient to express the annihilation operators $\hat{b}_k$ associated with the measurement modes $u_k(\vec{r})$ in terms of the annihilation operators $\hat{a}_\pm$ associated with the image plane modes $u_\pm (\vec{r})$: $\hat{b}_k = \sum_{j = \pm} f_{kj} \hat{a}_j$, with
\begin{align}
    f_{kj} = \int d \vec{r} u_k(\vec{r}) u_j (\vec{r}) = \frac{1 \pm (-1)^k}{\sqrt{2(1\pm\delta)}}\gamma_k,
    \label{eq:f}
\end{align}
with 
\begin{align}
    \gamma_k = \int d\vec{r} u_k(\vec{r}) u_0(\vec{r} + \vec{r}_1) = \frac{1}{\sqrt{k!}}e^{-s^2/8}\left(\frac{s}{2}\right)^{k}.
\end{align}
Accordingly, the mean photon number in the measurement modes is given by
\begin{align}
\label{eq:Nk}
    N_k &=  \langle \hat{b}^\dagger_k \hat{b}_k \rangle  = \sum_{ij=\pm} f_{ki}g_{kj} \langle \hat{a}^\dagger_i \hat{a}_j \rangle
    = |f_{k+}|^2 |\alpha_+|^2 + |f_{k-}|^2 |\alpha_-|^2 + 2\Re \left[ \alpha_+^* \alpha_-\right]f_{k+}f_{k-}  
\end{align}
where we used that the CARS signal is a coherent state $|\alpha_+,\alpha_-\rangle$.
As we did for direct imaging, we use the fact that the coherent states $|\alpha_+, \alpha_-\rangle$ have Poissonian photon number statistics to write the FI for SPADE as a simple function of mean photon number per mode  \cite{kay1993}
\begin{align}
    F_d^{\mathrm{SPADE}} = \sum_{k=0}^M \frac{1}{N_k}\left(\frac{\partial N_k}{\partial d} \right)^2.
    \label{eq:fi_spade_app}
\end{align}

\subsubsection{Plane-wave excitation pulses}
If we consider the excitation pulses to be plane waves, we can use Eqs.~\eqref{eq_app:alpha_pm_plane_wave} to express the mean photon number per mode as
\begin{align}
      N_k &= 2 g\kappa^2 \left[(1+\delta)|f_{k+}|^2 \cos^2\left(\frac{\tilde{k} s}{2}\right)  + (1-\delta)|f_{k-}|^2 \sin^2\left(\frac{\tilde{k} s}{2}\right)  \right] = 2g^2\kappa \gamma_k^2\left[1 + (-1)^k \cos ( \tilde{k} s) \right], 
\end{align}
where we used the expression for the overlaps \eqref{eq:f}. If we expand to leading order in $s$, for which terms with $k>2$ are irrelevant, we obtain 
\begin{align}
    \frac{w^2 F_d^{\mathrm{SPADE}}}{2\kappa g^2} \sim \left(3+6\tilde{k}^2+\tilde{k}^4\right) s^2,
\end{align}
that coincide with the QFI~\eqref{eq:QFI-small} in this regime. For $\tilde{k} =0$, we can take the sum for $M\to \infty$ analytically, and we obtain 
\begin{align}
    \frac{w^2 F^{\mathrm{SPADE}}_d}{2\kappa g^2} = 1 +e^{-s^2/2}(s^2-1),
\end{align}
which coincide with the QFI~\eqref{eq:QFI_plane} for all values of $s$. 
For $\tilde{k} \neq 0$, we couldn't evaluate the infinite sum, but analytical results for finite $M$ show how increasing $M$ the FI for SPADE approaches the QFI for larger and and larger separation (see Fig.~\ref{fig:fi-spade}).
\begin{figure}
    \centering
    \includegraphics[width=0.4\columnwidth]{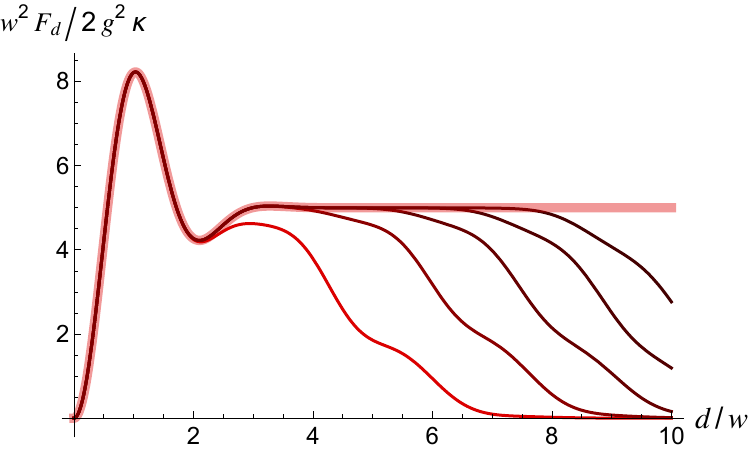}
    \caption{Fisher information for separation estimation for SPADE with $M = 5, 10, 15, 20, 25$ (lighter to darker curve). The broad red band correspond to the QFI. All curves were obtained with $\tilde{k} = 2$.}
    \label{fig:fi-spade}
\end{figure}

\subsubsection{Laguerre-Gauss Stokes pulse}
We now look at the case where the pump pulse is the plane wave $u_{\mathrm{pu}}(\vec{r}) = 1$ and the Stokes pulse is the LG mode $u_{\mathrm{St}}(\vec{r}) = \mathcal{N}_{\mathrm{St}} r \exp (i\phi) \exp (-r^2/w_{\mathrm{St}}^2)$. 
As we did before, we want to consider the possibility of the LG Stokes pulse is not aligned with the centroid of the two sources.
Nevertheless, we assume that the SPADE basis is centered on the sources centroid. 
Under these assumptions, we can compute the mean number of photons per mode substituting Eq.~\eqref{eq:alphas_LG_y} into Eq.~\eqref{eq:Nk}
\begin{equation}
    N_k = \frac{g^2 \kappa e}{a^2}\frac{e^{-s^2/2}e^{-(s^2/4a^2}e^{-2\psi^2/a^2}}{2 k!} \left(\frac{s}{2}\right)^{2k}\left[ \left(1 - (-1)^k\right)^2s^2 + 4\left(1 + (-1)^k\right)^2\psi^2 \right].
\end{equation}
For $M\to \infty$  it is in general not possible to take the sum in Eq.~\eqref{eq:fi_spade_app} analytically, however analytical results for finite $M$ shows that the FI for SPADE coincide with the QFI for all considered values of the separation.
If we set $\psi=0$, then we can take the infinite sum analytically, and we obtain 
\begin{equation}
\label{eq:FI-SPADE-lg}
    \frac{w^2 F^{\mathrm{SPADE}}_d}{2\kappa g^2}  = \frac{e}{2a^6}e^{-s^2/2a^2}\left[s^4 +a^2s^2(a^2-4) +4a^2 + e^{-s^2/2} \left( (a^2-1)^2s^4+a^2(5a^2-4)s^2 + 4a^4\right)\right],
\end{equation}
which coincides with the QFI~\eqref{eq:QFI-lg} in the main text.
\end{widetext}
\end{document}